\documentclass[fleqn,10pt]{elsarticle} \title{ Intrinsic basis-independent quantum coherence measure  }

\author[1]{Wei-Chen Wang} \author[2,*]{Mao-Fa Fang} \author[2]{Min Yu}
  \address[*]{Corresponding author. E-mail:mffang@hunnu.edu.cn}
\address[1]{Department of Physics, The National University of Defense Technology, Changsha 410073, Peoples Republic of China}
\address[2]{Key Laboratory of Low-Dimensional Quantum Structures and Quantum Control of Ministry of Education, and Department of Physics, Hunan Normal University, Changsha 410081, People's Republic of China}
\begin{document}
\begin{abstract}
\indent
 Quantum coherence is a key resource in quantum information processing scenarios, and quantifying coherence is an important task for both
 quantum foundation and quantum technology. However, until now, all most of coherence measures are basis-dependent that does not accord
 with physical reality, since the physical properties of the physical system should not be changed with the different choice of coordinate
 systems. Here, we propose an \textit{intrinsic basis-independent quantum coherence measure} which satisfies all conditions for quantifying
 coherence. This measurement not only reveals physical essence of quantum coherence of the quantum state itself clearly, but also
 simplifies the measurement procedure by avoiding the optimization procedure of distance measure.

\end{abstract}

\flushbottom \maketitle 

\section*{INTRODUCTION} \setlength{\arraycolsep}{0.1em}
 Quantum coherence (QC) originating from quantum superposition plays a central role in quantum mechanics,
     and is a key resource in quantum information processing scenarios such as quantum reference frames \cite{01,02,03}, transport in
     biological systems \cite{04,05,06} and quantum thermodynamics \cite{07,08,09}. How to measure QC  is an important problem in both
     quantum theory and quantum information science, and has attracted much attention recently \cite{10}. A  framework to quantify  QC  of
     quantum
    states in the resource theories has been recently proposed by Baumgratz et. al \cite {10}. By following this framework, a number of
QC  measures based on various physical contexts have been put forward.  Norm of coherence and the relative entropy of coherence were first suggested as two QC measures based on distance measures \cite{10}. The QC  measures based on entanglement \cite{11}, operation \cite{12a,13a} and convex-roof construction \cite{14a,15a} were subsequently proposed. With QC  measures, various properties of QC, such as the relations between QC and other quantum resources \cite{11,16f,16a}, the QC in infinite-dimensional systems \cite{16b,16c}, the complementarity relations of QC \cite{16d}, and the measure of macroscopic coherence\cite{16e}, have been discussed. The quantification of QC  promoted in a resource-theoretic framework thus encouraged many further considerations about QC \cite{17a,17b,17c}.

   We would like to point out that, until now,  most of these studies on quantifying QC are not very perfect and are not applicable in
   certain situations. For example, the quantification of QC depends on a particular reference basis, namely, these QC measures are
   basis-dependent. As we known, the physical laws and properties of physical systems do not change with the different choice of reference
   basis. Therefore, QC should also be a physical property of the quantum state itself in quantum  system.  Previous basis-dependent QC
   measures cannot reflect physical reality related to the quantum state itself in quantum system. In this paper, we try to solve this
   problem, i.e., to establish a basis-independent QC  measure. By refining an accurate and unique zero-coherence state as the complete
   incoherent state, we propose a new QC  measure, called as an \textit{intrinsic basis-independent quantum coherence measure} (IBIQCM),
   which not only satisfies all the conditions in the resource-theoretic framework of quantifying QC, but also is basis-independent and can
   reflects the physical essence of the quantum state itself. Moreover, the new QC measure also simplifies the QC calculation.

\section*{Results and Discussion} \subsection*{Basis-dependent QC measures} \setlength{\arraycolsep}{0.1em}
  First we review some previous basis-dependent approaches for QC measures  \cite{10}. Fixing a reference basis in a $d$-dimensional Hilbert
  space $\mathcal{H}$, say $\{ |i\rangle \}_{i=0}^{d-1}$, any quantum state can be expressed by $\rho=\sum_{i,j}\rho_{i,j}|i\rangle\langle
  j|$. Label the set of incoherent states by $\mathcal{I} \subset \mathcal{H}$, where the incoherent states are diagonal states in the
  reference basis, i.e.,
\begin{equation} \delta =\sum_{i=1}^d \delta_i |i\rangle\langle i| \label{incostate} \end{equation} The incoherence operators are defined to fulfil $K_n \mathcal{I} K_n^{\dagger} \subset \mathcal{I}$ for all $n$, where quantum operations are specified by a set of Kraus operators $\{ K_n\}$ satisfying $\sum_n K_n  K_n^{\dagger} = \mathbf{1}$. Following the approach to entanglement measurement, any proper measure of coherence for the state $\rho$ requires a set of conditions:

\begin{description} \item[(C1)] $C(\delta) = 0$ iff $\rho$ belongs to the set of incoherent states $\mathcal{I}$ (i.e., $\delta \subset \mathcal{I}$). \item[(C2)] Monotonicity under incoherent completely positive and trace preserving maps $\Phi_{ICPTP} = \sum_{n} K_n \rho K_n^{\dagger}$,
    i.e.,  $C(\rho) \ge C(\Phi_{ICPTP} (\rho)$, and under selective measurements on average, i.e., $C(\rho) \ge \sum_n p_n C(\rho_n)$,
    where the retained state corresponding to outcome $n$ after a measurement $\rho_n =  K_n \rho K_n^{\dagger} /p_n$ with probability $p_n
    = tr[K_n \rho K_n^{\dagger} ]$.
\item[(C3)] Nonincreasing under missing of quantum states (convexity), i.e., $\sum_n p_n C(\rho_n) \ge C(\sum_n p_n\rho_n)$ for any set of
    states $\rho_n$ and any $p_n \ge 0$ with $\sum_n p_n = 1$.
\end{description}

Several quantities have been proposed as candidates for measuring QC \cite{10,11,12a,13a,14a,15a}. For example, the candidate coherence measurement can be defined according to the distance measure $\mathcal{D}$ between quantum states
 \begin{equation} C_{\mathcal{D}}(\rho) = \min_{\delta \in \mathcal{I}} \mathcal{D}(\rho, \delta) \label{C_D} \end{equation}
 Therefore, two proper measures of QC \cite{10} are the $l_1$ norm of coherence
    \begin{equation}\label{06}
    C_{n}(\rho)=\sum_{i,j(i\neq j)}  |\rho_{i,j}|.
    \end{equation} 
and the relative entropy of coherence which describe coherence of quantum states:
    \begin{equation}\label{07}
    C_{re}(\rho)=S(\rho_{diag})-S(\rho),
   \end{equation}
    where $S(\rho) = -tr(\rho \log_2 \rho)$ is Von Neumann entropy, and $\rho_{diag}=\sum_{i}\rho_{i,i}|i\rangle\langle i|$ denotes the
diagonal part of $\rho$. It should be pointed out that a striking feature of the relative entropy of coherence is that it  connectes the QC with the purity of quantum states via von Neunnman entropy. From the inequation $C_{re}(\rho)\leq S(\rho_{diag})\leq \log(d)$ and the definition of relative entropy of coherence, it can be see that, the smaller $S(\rho)$, the higher purity of quantum states, the stronger $C_{re}(\rho)$. 

Although these measures satisfy the above conditions (C1),(C2) and (C3),  all they are defined in a particular reference basis. Therefore, for the same quantum state $\rho$,  by choosing different reference basis, the results obtained by these QC  measures are completely different.  For example, for the diagonal state $\varrho_z$ of a qubit in Pauli operator $\sigma_z$ basis
    \begin{eqnarray}\label{03}
     \varrho_z=\left(
       \begin{array}{ccc}
         \cos^{2}\alpha & 0  \\
         0 & \sin^{2}\alpha \\
       \end{array}
     \right),
    \end{eqnarray}
both the relative entropy of coherence $ C_{re}(\varrho_z)$ and the norm of
    coherence $ C_n(\varrho_z)$ are zero because all off-diagonal elements of the density matrix $\varrho_z$ are zero.
But if $\varrho_z$ is written down in $\sigma_x$ basis, the density matrix will be
    \begin{eqnarray}\label{04}
     & & \varrho_x= H\varrho_{z}H=\frac{1}{2} \left(
       \begin{array}{ccc}
         1 & \cos(2\alpha)  \\
         \cos(2\alpha) & 1\\
       \end{array}
     \right),  \\
    & & H=\frac{1}{\sqrt{2}}\left(
       \begin{array}{ccc}
          1 & 1  \\
         1 & -1\\
       \end{array}
     \right), \nonumber
     \end{eqnarray}
  where $H$ is Hadamard gate, which is unitary operator. When $\alpha$ is not $\frac{\pi}{4}$ or $\frac{3\pi}{4}$, the off-diagonal elements
  of $\varrho_x$ are nonzero. Then, both $ C_{re}(\varrho_x)$ and $C_{n}(\varrho_x)$ are also nonzero. It is obvious that, the
    two QC  measures depend on the chosen reference basis. Meanwhile, physically, basis-dependent QC measures can also lead to a conclusion
    that does not accord with physical reality: according to the conclusion obtained by basis-dependent QC measures, before the quantum state
    of quantum system is measured, one can change the measuring basis to affect the QC  of quantum state itself. Therefore,
    \emph{Basis-dependent QC measures cannot reflect the intrinsic QC of the quantum state itself} and are not good QC measures. A good QC
    measure, thus, requires basis-independent to overcome the shortcoming.

      In order to quantify QC  precisely, it is important to exactly define the zero-coherence state, namely, we must find a complete
      incoherent state as the zero-coherence state. In basis-dependent QC measures, the incoherent states are defined as
      Eq.(\ref{incostate}), i.e., the coherence of all the states $\delta$ is zero. But we find that \emph{the definition of incoherent
      states in basis-dependent QC measures is not accurate.} In fact, the states defined by Eq.(\ref{incostate}) are not complete incoherent
      states, but partial incoherent states, except for all $\delta_i$ take the same constant.  This stems from the fact that, the QC  of the
      quantum state cannot be determined uniquely by the off-diagonal elements of the density matrix, which means that, there may be a
      certain QC  for a quantum state described by the density matrix with the zero off-diagonal elements. If we choose the state defined by
      Eq.(\ref{incostate}) as zero-coherence state, this will lead to a result that zero-coherence state depend on the choice of the
      reference basis, that is, the change of the reference basis will change their zero-coherence property. Therefore, in order to find an
      intrinsic basis-independent QC measure, one must re-define the zero-coherence state which is independent of the reference basis as the
      complete incoherent state.
\subsection*{Defining the zero-coherence state}
  \setlength{\arraycolsep}{0.1em}
   Since the feature of QC is impossible within classical physics, we must consider a maximum classical mixture state as the accurate and
   unique zero-coherence state.  According to the density matrix expression method  of quantum state, a  maximum classical mixture state can
   be described by the density matrix  with equal diagonal elements (equal probability) and zero off-diagonal elements. Thus we define the
   maximum classical mixed state $\delta_0$ as the $complete$ $incoherent$ state
    \begin{equation}\label{14}
    \delta_0=\frac{1}{d}\sum_{i=1}^{d}|i\rangle\langle i|,
    \end{equation}
which is basis-independent and does not possess any QC in $d$-dimensional Hilbert space. Obviously, the polarization state of natural light, which is incoherence light in experiment of interference of polarized light, can be described by the maximum classical mixed state.

 Given a $d$-dimensional Hilbert space, there is only one $complete$ $incoherent$ state $\delta_0$, which  is different from the $incoherent$
 state defined  by Eq.(\ref{incostate}). It is easy to prove the $complete$ $incoherent$ state $\delta_0$ is invariant under arbitrary
 unitary transformation, which means that, there is not any QC in the $complete$ $incoherent$ state in arbitrary basis. Seeing that the
 $complete$ $incoherent$ state $\delta_0$ possesses both the properties of the uniqueness and the basis-independence, an accurate and unique
 \emph{zero-coherence} state can be determined by it. Obviously, for arbitrary completely positive and trance preserving map $\Phi_{ICPTP}$,
 $\delta_{0}=\Phi_{ICPTP}(\delta_{0})$ is always satisfied.

 \subsection*{Basis-independent QC measures: IBIQCM}
 After introducing the accurate and unique  \emph{zero-coherence} state $\delta_0$, we can define a new basis-independent QC measure by the
 distance measure Eq. (\ref{C_D}), i.e., by comparing the distinction between a quantum state $\rho$ and the zero-coherence state $\delta_0$
 to quantify QC. Here we still use the relative entropy \cite{17} for the distance measure. Therefore, this new QC measure also describes the coherence of quantum states by using purity of quantum states and is defined by
      \begin{eqnarray}\label{08}
   C_{IBIQC}(\rho)&=&-tr[\rho\log\delta_{0}]+tr[\rho\log\rho], \nonumber \\
   &=&-tr[\rho\log(\frac{1}{d}\sum_{i=1}^{d}|i\rangle\langle i|)]+tr[\rho\log\rho] \nonumber \\
   &=& \log d-S(\rho)
    \end{eqnarray}
    by substituting $\delta_0$ given by Eq.(\ref{14}), where $S(\rho)$ is still Von Neumann entropy.  Equation (\ref{08}) is the central
    result of this paper.
    We call this QC  measure as $intrinsic$ $basis$-$independent$ $quantum$ $coherence$ $measure$ (IBIQCM) and denote it by
    $C_{IBIQC}(\rho)$.

According to the central result, we make the following discussion:

(1) \emph{IBIQCM is a kind of explicit, succinct and basis-independent coherence measure.}

It is obvious from Eq.(\ref{08}) that, although IBIQCM relates to  the purity of quantum state via the Von Neumann entropy as same as  the relative entropy of coherence, it is independent on the choice of the reference basis due to the first term $ \log d$ in Eq.(\ref{08}) only depend on the dimension of Hilbert space. Obviously, the basic characteristics of IBIQCM is able to be obtained from basic the quantum entropy\cite{17aa}. Moreover, since IBIQCM are basis-independent, arbitrary unitary operators are strictly incoherence operators in our scheme. However, we should note that the unitary operator can be either incoherence operator or coherence operator for different quantum states in framework of basis-dependent QC measures, for instance, Hadamard gate $H$ is coherence operator for $\rho_z$, but it is incoherence operator for $\delta_0$. Thus categorizations of incoherence operators and coherence operators are ambiguous in framework of basis-dependent QC measures.

In IBIQCM, since there is only one $complete$ $incoherent$ state  in $d$-dimensional quantum system, the optimization procedure  of distance measurement $\mathcal{D}(\rho,\delta)$ can be avoided, which leads to a great simplification for the  calculation of QC measure.

By the similar procedure to the relative entropy of coherence $C_{re}(\rho) $ \cite{10},  one can prove generally that $C_{IBIQC}(\rho)$ satisfies all the conditions (C1) (C2) and (C3) in the resource-theoretic framework of quantifying coherence, and an additional condition \begin{description} \item[(C0)] $C(\rho)$ is invariant under unitary transformations, namely $C(\rho)$ is basis-independent. \end{description}

(2) \emph{IBIQCM is intrinsic QC  measure reflecting the physical essence of QC of the quantum state itself.}

 Considering that Von Neumann entropy $S(\rho)$ is quantum information entropy about quantum systems \cite{08}, it is apparent from
 Eq.(\ref{08}) that $C_{IBIQC}(\rho)$ is a measure of the information of quantum state $\rho$ which we have obtained.
  If  quantum system  is in a pure state, we would have zero Von Neumann entropy and the maximum $C_{IBIQC}(\rho) = \log d$, namely we can
  obtain all quantum information from system without any losing in a certain time $t$, and the corresponding state is just a complete
  coherence state. When quantum system is in the maximum classical mixture state which loses all quantum information about system, we would
  have the maximum Von Neumann entropy $\log d$ and zero $C_{IBIQC}(\rho)$, namely we know nothing about system in a certain $t$, and the
  corresponding state is just the complete incoherent state.

   In order to further  examine the validity of IBIQCM, we return to the problem of $\varrho_z$ and $\varrho_x$ mentioned above. Substituting
   $\varrho_z$ and $\varrho_x$ into Eq.(\ref{08}), one obtains
   \begin{eqnarray}\label{09}
    C_{IBIQC}(\varrho_z)
   &=&\log 2+\cos^2\alpha\log\cos^2\alpha\nonumber\\
   &+&\sin^2\alpha\log\sin^2\alpha,\nonumber \\
   C_{IBIQC}(\varrho_x)&=&\log 2+\cos^2\alpha\log\cos^2\alpha\nonumber\\
    &+&\sin^2\alpha\log\sin^2\alpha.
  \end{eqnarray}
  Obviously, $C_{IBIQC}(\varrho_z)$ is equal to $C_{IBIQC}(\varrho_x)$, namely, quantum states $\varrho_z$ and $\varrho_x$ possess the same
  QC. Therefore, IBIQCM reflects the intrinsic property of the quantum state according with physical reality whatever the reference basis is.
  However, substituting $\varrho_z$ and $\varrho_x$ into Eq.(\ref{07}), one gets
  \begin{eqnarray}\label{10}
    C_{re}(\varrho_z)&=0 \nonumber\\
    C_{re}(\varrho_x)&=&\log 2+\cos^2\alpha\log\cos^2\alpha\nonumber\\
    &+& \sin^2\alpha\log\sin^2\alpha.
  \end{eqnarray}
$C_{re}(\varrho_x)$ is not equal to $ C_{re}(\varrho_z)$, which means that the relative entropy of coherence is basis-dependent.

 In addition, from Eqs. (\ref{09}) and  (\ref{10}) , it can be seen that in $\sigma_x$ basis, the results are the same for both the
 IBIQCM and the relative entropy of coherence, which means that the correct result can be obtained for the relative entropy of coherence only
 under some particular reference basis.

 In order to further illustrate the differences and connections between IBIQCM and the relative entropy of coherence, we design a polarized-light interference experiment device composed of a crystal wave plate with a phase delay and a linear polarizer.

 Here, we discuss two cases: Firstly, let a coherent plane polarized state project to vertically the crystal wave plate in Angle of $\frac{\pi}{4}$ between its vibration direction and the principal section of crystal wave plate (corresponding to a base vector selection).  After passing the crystal wave plate, the coherent plane polarized state is decomposed into two mutually vertical vibration components with the fixed phase relationship. Then, the two components project to the direction of the transmission light of the polarizer, and generate two mutually parallel vibration coherent components which can produce interference phenomenon behind the polarizer. If let a coherent plane polarized state pass the crystal wave plate in which the vibration direction is parallel to the direction of the principal section of the crystal wave plate, (corresponding to another base vector selection), there is only one coherent component behind crystal wave plate. After this coherent component passing the polarizer, there is still one coherent component and interference phenomenon cannot be produced. It is obvious that, the generation of the interference phenomenon requires not only the coherent source, but also the reasonable interference experiment design (corresponding to a reasonable base vector selection). If the interference experiment is not designed correctly (corresponding to the base vector selection is incorrect), even if there is coherent source, interference phenomenon can not be still produced; Second, let an incoherent nature light project to vertically the crystal wave plate, behind the crystal wave plate, although two mutually vertical vibration components can be produced, there is not fixed phase relationship between the two vibration components. Through the polarizer projection, the two mutually parallel vibration components naturally cannot produce any interference phenomenon.

 The above discussion shows that, the coherence of coherence source is independent on the design of interference experiment device, namely the coherence of quantum state itself is independent on base vector selection, while generation of interference phenomenon is dependent on the both the coherence of coherence source and design of interference experiment, namely, generation of interference phenomenon is dependent on the both the coherence of quantum state itself and experiment base vector selection.

 IBIQCM is an essential measure and description for the coherence of the quantum state itself, but it cannot provide judgment on whether the interference phenomenon can occur. While the relative entropy of coherence can provide a judgment for the possibility of experimental interference phenomenon occur if the coherent source is given, but it cannot provide the essential measurement and description for the coherence of quantum state itself.

On the other hand, A maximally coherent state, which is defined by literature\cite{10}, can be written as
   \begin{eqnarray}\label{11}
    |\Psi\rangle=\frac{1}{\sqrt{d}}\sum_{i=1}^{d}|i\rangle.
  \end{eqnarray}
  According to the theory of coherence of electromagnetic
  field\cite{18b}, coherence state $|\alpha\rangle=exp(-\frac{1}{2}|\alpha|^{2})\sum_{n=0}^{\infty}\frac{\alpha^{n}}{\sqrt{n!}}|n\rangle$ is
  totally coherent, but it is not maximally coherent state obviously in accordance with (\ref{11}). However in our scheme, coherence state
  $|\alpha\rangle$ is a maximally coherent state, thus we have reason to believe that our scheme is better than previous schemes.

\section*{Conclusions}
 In summary, we have proposed a new intrinsic basis-indpendent QC measure (IBIQCM) which reflects physical reality of the quantum state
 itself. More in details,  we have discussed some defects in previous basis-dependent QC measures, such as that basis-dependent QC measures
 cannot reflect the intrinsic QC of the quantum state itself, and the definition of the zero-coherence states is not accurate. For the
 resource-theoretic framework of quantifying coherence, in order to find better QC measures, an additional condition that intrinsic QC
 measures must be basis-independent should be supplemented. By redefining
  the maximum classical mixture state as a complete incoherent state, i.e., it is the accurate and unique zero-coherence state for the finite
  dimensional Hilbert space,  we have presented an IBIQCM, which can reflect the intrinsic QC property of the quantum state itself, by using
  the relative entropy. The IBIQCM satisfies all the conditions of the resource-theoretic framework of quantifying coherence. Moreover, the
  new IBIQCM reveals physical essence of quantum coherence, and simplifies the calculation of QC measure. Based on the importance of
  quantifying coherence in QC investigations, our research result is an important progress.

\section*{Outlook} First, in definition of  IBIQCM,  since  the relative entropy  plays a role of  distance measures between any one quantum state and the complete incoherent state,  there are other potential candidates for distance measures, which we have not discussed here, such as fidelity, trace norm \cite{18a} and so on.
  one can also define many other QC measures  similar to IBIQCM by using these potential candidates. Secondly, it is  meaningful for
  studying or restudying  various properties of quantum coherence by using IBIQCM,
such as the relations between quantum coherence and other quantum resources \cite{11,16f,16a}, the complementarity relations of quantum coherence \cite{16d}, cohering, decohering power of quantum channels \cite{19} and so on. Thirdly, it is important for generalizing IBIQCM to situation of infinite dimensional systems \cite{16b,16c}, because of many quantum states and quantum physical systems are discussed in infinite dimensional systems necessarily. Finally, in our scheme, owing to the reason that zero-coherence state is unique in $d$-dimensional Hilbert space, the optimization procedure of distance measure can be avoided. Therefore, it is valuable  to consider that whether other quantum resource theory, such as entanglement and quantum discord, has this advantage that can greatly simplify the calculation.

\section*{References}
\bibliography{sample}

\bibliographystyle{model1a-num-names}
 \section*{Author Contributions}
 W.W. contributed the idea. W.W. performed the calculations. M.F. checked the calculations. W.W.
wrote the main manuscript, M.F. made an improvement. All authors contributed to discussion and reviewed the manuscript. \section*{Acknowledgements}

 \setlength{\arraycolsep}{0.1em}
We thank professor Xin-Hua Peng for her helpful discussions. This work was supported by the National Natural Science Foundation of China (Grant Nos.11374096) and Hunan Provincial Innovation Foundation for Postgraduate (CX2017B177). \section*{Additional Information} Competing financial interests: The authors declare no competing financial interests \end{document}